\documentclass[preprint,prd,showpacs]{revtex4}
\usepackage{epsfig}
\begin{document}
\title{
Neutrino self-energy in external magnetic field 
      }
\author{Andrea Erdas}
\affiliation{
 Department of Physics, Loyola University Maryland, 4501 N Charles St
       Baltimore, MD 21210}
\email{aerdas@loyola.edu}
\begin {abstract} 
Using the exact propagators
in a constant magnetic field, the neutrino self-energy has been calculated
to all orders in the field strength $B$ within the minimal extension of the Weinberg-Salam model
with massive Dirac neutrinos. 
A simple and very accurate formula for the self-energy is obtained, that is valid for
$0\le B\ll m^2_W/e$ and for neutrino transverse momentum to the magnetic field 
$p_{\perp}\ll m_W$. I discuss the implications of this finding 
to the dispersion of massless neutrinos in vacuum and in a
charge-symmetric medium, and to the magnetic field induced resonance transitions of massive 
neutrinos inside supernovae and magnetars,
and calculate the neutrino magnetic moment. 
\end {abstract}
\maketitle
\section{Introduction}
The study of creation, propagation, energy loss, and absorption of
neutrinos in magnetic field is important in many astrophysical
contexts and in early cosmology~\cite{raffeltbook}. The neutrino
self-energy and dispersion relation are modified in magnetized
media and in a vacuum with magnetic field, and such modifications have been studied extensively in the 
literature ~\cite{erdaskim,dolivo,elmfors,erdasfeld,kuznetsov2,mckeon,elizalde1,elizalde2}.
There is a  natural scale for the magnetic field strength required to significantly impact
quantum processes, and it can be expressed in terms of the electron mass $m_e$ and the elementary charge $e$ as 
$B_e=m^2_e/e\simeq 4.41 \times 10^{13}$ G. Large magnetic fields are present in a variety of 
astrophysical sites like supernovae, neutron stars and white dwarfs, and fields as large as
$B_e$ or larger can arise in supernovae explosions or coalescing neutron stars. The remnants of such astrophysical cataclysms
are magnetars, young neutron stars with magnetic fields $10^{14}-10^{16}$ G \cite{Duncan,Thompson1,Thompson2}.
It has been suggested that during the electroweak
phase transition local magnetic fields much stronger than those of a magnetar could have existed, with field strength
as high as
$10^{22}-10^{24}$ G \cite{Brandenburg,Joyce,grasso}. Situations where even
stronger magnetic fields could exist in extreme astrophysical
and cosmological environments are possible.
While neutrinos might rarely encounter magnetic fields larger than $10^{16}$ G, many 
situations arise where an abundant production of neutrinos occurs in astrophysical sites such as supernovae, 
neutron stars, white dwarfs and magnetars where magnetic field
strengths can be at or around $B_e$. A literature search reveals that, while calculations 
of the neutrino self-energy in magnetic field
have a long history \cite{erdasfeld,kuznetsov2,mckeon,elizalde1,elizalde2}, only in the last few years
has this matter been partially settled 
with the paper by Kuznetsov et al.\cite{kuznetsov2}
who, using an expansion for the $W$-propagator where only the lowest order terms are retained,
find the correct asymptotic values of the self-energy for neutrino transverse momentum $p_{\perp}\ll m_W$ 
in the case of weak magnetic field $eB\ll m^2_e$ and 
of moderate magnetic field $m^2_e\ll eB\ll m^2_W$. 
We still do not know the self-energy or dispersion relation of neutrinos with low transverse momentum for 
magnetic field strengths that are not much smaller and not much bigger than $B_e$, 
and therefore a calculation of the neutrino self-energy
and dispersion relation for $p_{\perp}\ll m_W$ that is valid for magnetic field strengths covering the 
whole range $0~\le~eB\ll~m^2_W$ is needed.

In  this paper I use Schwinger's proper time method \cite{schwinger} to
calculate the neutrino self-energy in homogeneous magnetic fields within the minimally extended Standard Model 
of electroweak interactions with massive Dirac neutrinos. Using the 
exact $W$-propagator I obtain a simple and very accurate analytic form of the self-energy 
that is valid for $0~\le~eB\ll~m^2_W$ and when the neutrino transverse momentum to the magnetic field is $p_{\perp}\ll m_W$, and I show that, 
in the case of weak and moderate fields, my result 
agrees with the results of Ref. \cite{kuznetsov2}. I discuss the implications of this finding 
to the dispersion of massless and massive neutrinos in a plasma and in a vacuum with magnetic field 
and use it to calculate the neutrino magnetic moment.

In Section \ref{2} the notation for the fermion, gauge boson and scalar propagators in magnetic field \cite{erdasfeld} is reviewed
and the one-loop neutrino self-energy is set up in the framework of the minimal extension of the Standard Model 
\cite{kuznetsov2,kuznetsov3}. 
In Section \ref{3} I calculate the self-energy and obtain simple analytic expressions for all its terms.
An extended discussion of the implications of my results and the conclusions are in Section \ref{4}.
\section{ Propagators and neutrino self-energy in a constant magnetic field}
\label{2}
The metric used in this paper is
$g^{\mu \nu} = \mathrm{diag}(+1,-1,-1,-1)$ and the $z$-axis points in the direction of the constant 
magnetic field $\mathbf{B}$. Therefore the electromagnetic
field strength tensor $F^{\mu \nu}$ has only two non-vanishing components $F^{21}= -F^{12} = B$. 

For the purpose of this work, it would seem convenient to work in the
unitary gauge where the unphysical scalars
disappear. 
However, the $W$-propagator is quite cumbersome in this gauge, and I prefer
to work in the Feynman gauge, where the $W$-propagator has a much simpler expression. The
following expressions for the charged lepton
$S(x',x'')$ \cite{schwinger,dittrich}, $W$-boson
$G^{\mu \nu}(x',x'')$ and scalar propagators $D(x',x'')$  \cite{erdasfeld} in a constant
magnetic field have been written using Schwinger's proper time method:
\begin{equation}
S(x',x'')=\Omega(x',x'')\!\int{d^4k\over (2\pi)^4}e^{-ik\cdot (x'-x'')}
S(k) \quad , 
\label{sx}
\end{equation}
\begin{equation}
G^{\mu\nu}(x',x'')=\Omega(x',x'')\!\int{d^4k\over (2\pi)^4}e^{-ik\cdot
(x'-x'')}
G^{\mu\nu}(k) \quad ,
\label{gx}
\end{equation}
\begin{equation}
D(x',x'')=\Omega(x',x'')\!\int{d^4k\over (2\pi)^4}e^{-ik\cdot (x'-x'')}
D(k) \quad ,
\label{dx}
\end{equation}
and, in the Feynman gauge, the translationally 
invariant parts of the propagators are
\begin{equation}
\label{S}
S(k) =\int_0^\infty \!\!{ds\over\cos eBs} 
{\exp{\left[-is\left(m^2_\ell-k^2_{\parallel}-
k^2_{\perp}{\tan eBs\over eBs}\right)\right]}}
\left[(m_\ell+
\not\! k_{\parallel})e^{-ieBs\sigma_3}+
{\not\! k_{\perp}\over \cos eBs}
\right] , 
\end{equation}
\begin{equation}
G^{\mu \nu}(k)=
-\int_0^\infty \!\!{ds\over\cos eBs}\,
{\exp{\left[-is\left(m^2_W-k^2_{\parallel}-
k^2_{\perp}{\tan eBs\over eBs}\right)\right]}}
[g^{\mu \nu}_{\parallel}
-(e^{2eFs})^{\mu\nu}],
\label{G}
\end{equation}
\begin{equation}
D(k)=
\int_0^\infty \!\!{ds\over\cos eBs}\,
{\exp{\left[-is\left(m^2_W-k^2_{\parallel}-
k^2_{\perp}{\tan eBs\over eBs}\right)\right]}},
\label{D}
\end{equation}
where $-e$ and
$m_\ell$ are the charge and mass of the charged lepton $\ell$, and
$m_W$ is the $W$-mass. It is convenient to use the notation
\begin{equation}
a^\mu_{\parallel}=(a^0,0,0,a^3), \quad
a^\mu_{\perp}=(0,a^1,a^2,0)
\label{amu}
\end{equation}
and 
\begin{equation}(a b)_\parallel=a^0\,b^0-a^3\,b^3, \quad
(a b)_\perp=-a^1\,b^1-a^2\,b^2
\label{adotb}
\end{equation}
for arbitrary four-vectors $a$ and $b$. Using this notation I write the metric tensor as
\begin{equation}
g^{\mu\nu}=g^{\mu\nu}_\parallel+g^{\mu\nu}_\perp
\label{g}
\end{equation} 
with 
\begin{equation}
g^{\mu\nu}_\parallel=\tilde{\varphi}^{\mu\alpha}\tilde{\varphi}_{\alpha}^{\,\,\,\nu}, \quad
g^{\mu\nu}_\perp=-\varphi^{\mu\alpha}\varphi_{\alpha}^{\,\,\,\nu},
\label{gpar}
\end{equation} 
 where $\varphi$ is 
the dimensionless electromagnetic field tensor normalized to $B$ and 
$\tilde\varphi$ is its dual~\cite{kuznetsov2}
\begin{equation}
{\varphi}^{\mu\nu}={F^{\mu\nu}\over B}, \quad
\tilde{\varphi}^{\mu\nu}={1\over2}\epsilon^{\mu\nu\alpha\beta}\varphi_{\alpha\beta} \quad . 
\label{varphi}
\end{equation} 
The $4\times 4$ matrix $\sigma_3$ that appears in the charged lepton propagator (\ref{S}), 
can be written in terms of $\varphi$ as
\begin{equation} 
\sigma_3=
{i \over 2}[\gamma^1, \gamma^2]=-{i \over 2}( \gamma{\varphi}\gamma ) \quad ,
\label{sigma3}
\end{equation} 
where the Lorentz indices of vectors and tensors within parentheses are contracted, e.g. 
$( \gamma{\varphi}\gamma )=\gamma_\mu{\varphi}^{\mu\nu}\gamma_\nu$. When writing the $W$-propagator (\ref{G}), I use the notation
\begin{equation}
\left(e^{2eFs}\right)^{{\mu}\nu}=
-g^{\mu\nu}_{\perp} \cos{(2eBs)}
+\varphi^{\mu\nu}  \sin{(2eBs)} \quad .
\label{efmunu}\end{equation}
I choose the electromagnetic vector potential to be
$A_\mu=-{1\over2}F_{\mu \nu}x^\nu$ and therefore
the phase factor which appears in Eqs. (\ref{sx}), (\ref{gx}), (\ref{dx}) is given by
\cite{dittrich}
\begin{equation}
\Omega(x',x'')=\exp\left(
-i {e\over 2}x'_\mu F^{\mu \nu} x''_\nu
\right)  \quad .
\label{2_08a}
\end{equation}

The neutrino self-energy operator $\Sigma(p)$ is defined in terms of the 
invariant amplitude for the transition $\nu_\ell\rightarrow~\nu_\ell$
\begin{equation}
{\cal{M}}(\nu_\ell\rightarrow\nu_\ell)=-\bar{\nu}(p)\Sigma(p)\nu(p).
\label{M}
\end{equation}
Perturbatively, the self-energy operator in the Feynman gauge corresponds to the sum of two 
diagrams, a bubble diagram with the gauge boson and a bubble diagram with the scalar
\begin{equation}
\Sigma(p)=\Sigma_W(p)+\Sigma_\Phi(p).
\label{Sigma}
\end{equation}
The translationally non-invariant phase factors $\Omega(x',x'')$ are identical for all propagators
and the product of phase factors in the two-vertex loop is
\begin{equation}
\Omega(x',x'')\Omega(x'',x')=1
\label{phiprod}
\end{equation}
therefore, within the minimally extended version of the standard model of
electroweak interactions with an $SU(2)$-singlet right-handed neutrino, the two bubble diagrams can be written as \cite{erdasfeld,kuznetsov3}
\begin{equation}
\Sigma_W(p)=-i{g^2\over 2}R{\gamma}_{\alpha}
\int{d^4k\over (2\pi)^4}S(p-k)G^{\beta\alpha}(k){\gamma}_{\beta}L,
\label{sigmaw}
\end{equation}
\begin{equation}
\Sigma_\Phi(p)=-i{g^2\over 2m^2_W}[m_\ell R-m_\nu L]
\int{d^4k\over (2\pi)^4}S(p-k)D(k)[m_\ell L-m_\nu R],
\label{sigmaphi}
\end{equation}
where $g$ is the $SU(2)$ coupling constant, $L={1\over 2}(1-\gamma_5)$ and
$R={1\over 2}(1+\gamma_5)$ are the left-handed and right-handed projectors
and neutrino mixing is allowed by taking a nondiagonal neutrino mass matrix
$m_\nu$ in Eq. (\ref{sigmaphi}).
\section{ Calculation of the self-energy}
\label{3}

Inserting the expression for the propagators from Eqs. (\ref{S}) and (\ref{G}) 
into the self-energy, I write ${\Sigma}_{W}(p)$ as
\begin{eqnarray}
\Sigma_{W}(p)=&&{ig^2\over 2}
\int\!{{d^4 k}\over{{(2{\pi})}^4}}{\int}^{\infty}_{0}
{ds_1\over{\cos z_1}}{\int}^{\infty}_{0}
{ds_2\over{\cos z_2}}{e^{-is_1(m^2_\ell -q^2_{\parallel}
-q^2_{\perp}{{\tan {z_1}}\over {z_1}})}}
{e^{-is_2(m_W^2 -k^2_{\parallel}-k^2_{\perp}{{\tan {z_2}}\over
{z_2}})}}{\times}
\nonumber \\
&&R{\gamma}_{\alpha}
\left[(m_\ell+
\not\! q_{\parallel})e^{-iz_1\sigma_3}+
{\not\! q_{\perp}\over \cos z_1}
\right]
[g^{\beta\alpha}_{\parallel}
-(e^{2eFs_2})^{\beta\alpha}_{\perp}]{\gamma}_{\beta}L
\end{eqnarray}
where
\begin{equation}
q = p-k \quad , \quad\quad z_1 = eBs_1 \quad , \quad\quad z_2 = eBs_2 \quad
.
\end{equation}
I do the straightforward $\gamma$-algebra, change variables from $s_i$
to
$z_i$, translate the $k$ variables
of integration as follows
\begin{equation}
(k_{\parallel}\,\,,\,\,k_{\perp})
\,\,{\rightarrow}\,\,
(k_{\parallel}+{z_1\over z_1+z_2}
p_{\parallel}\,\,,\,\,
k_{\perp}
+ {{\tan{z_1}}\over{{\tan{z_1}}+{\tan{z_2}}}}  p_{\perp})
\end{equation}
and, finally, perform the four gaussian integrals over the shifted variables
$k$. The result is:
\begin{eqnarray}
\Sigma_{W}(p)=&& -{g^2\over (4\pi)^2}
{\int}^{\infty}_{0}
{\int}^{\infty}_{0}
{dz_1 dz_2 \over (z_1+z_2) \sin(z_1+z_2)}
e^{-i[z_1m^2_\ell + z_2 m_W^2  - {\cal P}]/eB}  \times
\nonumber \\
&& \left[{z_2\over z_1+z_2}{\not\! p_{\parallel}}e^{
i\sigma_3(z_1+2 z_2)}+{\sin z_2 \over \sin(z_1+z_2)}
{\not\! p_{\perp}}\right]
\!L + (\textrm{c.t.})_W
\label{sigmaw2}
\end{eqnarray}
where
\begin{equation}
{\cal P}={z_1 z_2 \over (z_1 + z_2)}
p_{\parallel}^2
+{\sin{z_1} \sin{z_2} \over \sin{(z_1+z_2)}}
p_{\perp}^2 \quad ,
\end{equation}
and the appropriate counter-terms (c.t.) are defined such that
\begin{equation}
(\textrm{c.t.})_W=-{\Sigma}_W(p){\Bigr|}_{B=0,{\not\!\,p}=0}-{\not\! p}
{\biggl[}{{{\partial}{\Sigma}_W(p)}\over{{\partial}{\not\! p}}}
{\biggr]}_{B=0,{\not\!\, p}=0}.
\end{equation}
Next it is convenient to change integration
variables from $(z_1, z_2)$ to $(\tau, u)$ defined by
\begin{equation}
z_1 = \tau (1-u) \quad\quad \textrm{and} \quad\quad
z_2=\tau u \quad ,
\end{equation}
and to perform a clockwise rotation in the complex plane so that $\tau=-iz$. The result is
\begin{eqnarray}
\Sigma_{W}(p)&=& -{g^2\over (4\pi)^2}
{\int}^{\infty}_{0}{dz\over\sinh z}
{\int}^{1}_{0}du\,\,
e^{-(z/\eta)[(1-u)\lambda_\ell +  u+w(u,z)\xi]}  \times
\nonumber \\
&&R\left[u\cosh(z+zu){\not\! p_{\parallel}}+u\sinh(z+zu)\sigma_3{\not\! p_{\parallel}}
+{\sinh zu \over \sinh z}
{\not\! p_{\perp}}\right]
\!L + (\textrm{c.t.})_W
\label{sigmaw3}
\end{eqnarray}
where I use $p^2_{\parallel}\simeq -p^2_{\perp}$ and introduce the parameters 
$\eta=eB/m^2_W$, $\lambda_\ell=m^2_\ell/m^2_W$ and $\xi=p^2_{\perp}/m^2_W$ and the function
\begin{equation}
w(u,z)= u(1-u)-{\sinh zu \sinh (z-zu)\over z \sinh z}.
\end{equation}
At this point we must analyze the role played by the neutrino transverse momentum. A detailed analysis of the role
of $p_{\perp}^2$ is done in Ref. \cite{kuznetsov4} where the neutrino self-energy is calculated in 
terms of integrals of the Hardy-Stokes functions and it is found that the relevant dynamical field parameter is $\chi^2=\xi \eta^2$. 
The authors find that in the region of parameter values where $\chi$ is the smallest parameter 
in the problem, $\chi^2\ll\lambda_\ell$ or $eBp_{\perp}\ll m_\ell m^2_W$,
the self-energy has a dependence on $p_{\perp}$ only through a negligibly small imaginary part proportional to $e^{-\sqrt{3\lambda}/\chi}$.
A calculation of the contribution $\Sigma^n$ to the neutrino self-energy from the $n$th charged lepton Landau level \cite{kuznetsov2} 
(in conjunction with the exact $W$ propagator) also shows that, for low neutrino transverse momentum, $\Sigma^n$ does not
 depend on $p_{\perp}^2$, since $p_{\perp}^2$ only appears in the factor ${m^2_W\over
p_{\perp}^2}\ln\left(1+{p_{\perp}^2\over m^2_W}\right)$ which, for $p_\perp^2\ll m_W^2$, equals one. Therefore in the low transverse momentum limit
$p_\perp^2\ll m_W^2$, the self-energy does not depend on the parameter $\xi$ and one can neglect it from Eq.(\ref{sigmaw3}).
Since $B_W=m^2_W/e\simeq10^{24}$ G and $m_\ell
\ll m_W$, one can always
take $\eta\ll 1$ and $\lambda_\ell\ll 1$. After an integration
by parts of the factor in front of $\not\! p_{\perp}$, I obtain
\begin{eqnarray}
\Sigma_{W}(p)&=&-{g^2\over 16\pi^2}
{\int}^{\infty}_{0}{dz}
{\int}^{1}_{0}du\,\,
e^{-z\Lambda}\biggl[
(u\coth z\cosh zu-\Lambda\coth z\sinh zu+\Lambda u)
{\not\! p}L+
\nonumber \\
&+&
(\Lambda\coth z\sinh zu-\Lambda u+u\sinh zu)
{\not\! p_{\parallel}}L+\left.u{\sinh(z+zu)\over \sinh z}
\sigma_3{\not\! p_{\parallel}}L
\right]
\! + (\textrm{c.t.})_W
\label{sigmaw4}
\end{eqnarray}
where ${\not\! p_{\perp}}={\not\! p}-{\not\! p_{\parallel}}$ is used and
\begin{equation}
\Lambda={(1-u)\lambda_\ell + u\over\eta},
\end{equation}
\begin{equation}
(\textrm{c.t.})_W={g^2\over 16\pi^2}
{\int}^{\infty}_{0}{dz\over z}
{\int}^{1}_{0}du\,\,e^{-z\Lambda}u {\not\! p}L.
\end{equation}

I follow the same procedure to manipulate the expression of the bubble diagram with the
scalar, and obtain the following
\begin{eqnarray}
\Sigma_{\Phi}(p)&=&-{g^2\over 32\pi^2}
{\int}^{\infty}_{0}{dz}
{\int}^{1}_{0}du\,\,
e^{-z\Lambda} \biggl[
(u\coth z\cosh zu-\Lambda\coth z\sinh zu+\Lambda u)
{\not\! p}\,(\lambda_\ell L+\epsilon_\nu R)+
\nonumber \\
&&+(\Lambda\coth z\sinh zu-\Lambda u-u\sinh zu)
{\not\! p_{\parallel}}\,(\lambda_\ell L+\epsilon_\nu R)-\lambda_\ell m_\nu {\cosh(z-zu)\over \sinh z}
+
\nonumber \\
&&+ \lambda_\ell m_\nu{\sinh(z-zu)\over \sinh z} \sigma_3 -u{\sinh(z-zu)\over \sinh z}
\sigma_3{\not\! p_{\parallel}}\,(\lambda_\ell L+\epsilon_\nu R)
\biggr]
+(\textrm{c.t.})_\Phi
\label{sigmaphi2}
\end{eqnarray}
where I introduce the parameter $\epsilon_\nu=m^2_\nu/m^2_W$ and the counter-term is given by
\begin{equation}
(\textrm{c.t.})_\Phi={g^2\over 32\pi^2}
{\int}^{\infty}_{0}{dz\over z}
{\int}^{1}_{0}du\,\,e^{-z\Lambda}[u {\not\! p}\,(\lambda_\ell L+\epsilon_\nu R)-\lambda_\ell m_\nu].
\end{equation}

The self-energy operator I obtain, $\Sigma(p)=\Sigma_W(p)+\Sigma_\Phi(p)$, is valid for $p_{\perp}\ll m_W$ and
has the following Lorentz structure
\begin{equation}
\Sigma(p)=
\left[a_L {\not\! p} +b_L{\not\! p_{\parallel}}+c_L(p\,\tilde\varphi\gamma)\right]\!L+
\left[a_R {\not\! p} +b_R{\not\! p_{\parallel}}+c_R(p\,\tilde\varphi\gamma)\right]\!R
+m_\nu\left[K_1+iK_2(\gamma\varphi\gamma)\right]
\label{coefficients}
\end{equation}
since  
\begin{equation}
\sigma_3{\not\! p_{\parallel}}L=-(p\,\tilde\varphi\gamma)L,\;\;\;\;\;\;\;
\sigma_3{\not\! p_{\parallel}}R=(p\,\tilde\varphi\gamma)R
\end{equation}
and $\sigma_3=-{i \over 2}( \gamma{\varphi}\gamma )$. This Lorentz structure is in complete agreement with the findings
of Refs. \cite{kuznetsov2,kuznetsov3}. 

At this stage, we should discuss the meaning of the 
coefficients appearing in Eq. (\ref{coefficients}): 
$a_L$, $b_L$ and $c_L$ contain contributions from both diagrams, 
but the contribution from the diagram with the scalar is suppressed by a factor of $\lambda_\ell$.
The coefficients $a_L$, $a_R$ and $K_1$ are completely absorbed by the 
neutrino wave-function and mass renormalization. The $b$ and $c$ coefficients are relevant 
for neutrino dispersion but, to lowest order, the dispersion relation depends only on $b_L$ \cite{kuznetsov2}, 
thus the most relevant of these coefficients.
The two coefficients $b_R$ and $c_R$ 
are suppressed by a factor of $\epsilon_\nu$ relative to $b_L$ and $c_L$, and therefore play a less important role. 
However, in the case of a non-diagonal neutrino mass matrix, 
$b_R$ and $c_R$ might produce a modification of neutrino mixing 
in the presence of a magnetic field. 
The $K_2$ coefficient, along with $c_L$ and $c_R$,
is needed for the calculation of the 
neutrino magnetic moment \cite{kuznetsov3}. 

In the past, several authors have attempted to 
calculate $b_L$ \cite{erdasfeld,kuznetsov2,mckeon,elizalde1,elizalde2}, but all we know so far are its 
values in the limiting case of a "weak field" $eB\ll m^2_\ell$, and a "moderate field" $m^2_\ell\ll eB\ll m^2_W$.
Kuznetsov et al. \cite{kuznetsov2} found that $b_L={g^2\over 24\pi^2}\eta^2\left(\ln\lambda_\ell-{3\over4}\right)$ in the 
case of weak field and $b_L={g^2\over 24\pi^2}\eta^2\left(\ln\eta-2.542\right)$ for a moderate field. 
Their results are obtained using an expansion for the $W$-propagator where only terms up to second order 
in powers of the expansion parameter $eB$ are retained. In this paper $b_L$ and all the other
coefficients appearing in Eq. (\ref{coefficients}) are calculated using the 
exact $W$-propagator, and simple analytic forms for these coefficients will be obtained 
that are very accurate for $0\le eB\ll m^2_W$. It will be shown that, 
in the case of weak and moderate fields, the expression obtained here for $b_L$
agrees with the values obtained in Ref. \cite{kuznetsov2} in the appropriate limits. 

Eqs. (\ref{coefficients}) and (\ref{sigmaw4}) indicate that, to calculate $b_L$, we need to evaluate the following
double integral
\begin{equation}
b_L=-{g^2\over 16\pi^2}
{\int}^{\infty}_{0}{dz} {\int}^{1}_{0}du\,\,
e^{-z\Lambda}
(\Lambda\coth z\sinh zu-\Lambda u+u\sinh zu).
\label{bl2}
\end{equation}
We start by evaluating the last term
\begin{equation}
J(\eta,\lambda_\ell)=
{\int}^{\infty}_{0}{dz} {\int}^{1}_{0}du\,\,
e^{-z\Lambda} u\sinh zu.
\end{equation}
After an elementary but tedious integration we evaluate $J(\eta,\lambda_\ell)$ exactly and find
\begin{equation}
J(\eta,\lambda_\ell)=
{\eta^2\over{(1-\lambda_\ell)^2-\eta^2}}+{\eta\lambda_\ell\over 2(1+\eta-\lambda_\ell)^2}\ln\left({1+\eta\over\lambda_\ell}\right)
-{\eta\lambda_\ell\over 2(1-\eta-\lambda_\ell)^2}\ln\left({1-\eta\over\lambda_\ell}\right).
\end{equation}
Since $\eta\ll 1$ and $\lambda_\ell \ll 1$, we can expand in the two small parameters to obtain
\begin{equation}
J(\eta,\lambda_\ell)=
\eta^2[1+{\cal{O}}(\lambda_\ell\ln\lambda_\ell)]+{\cal{O}}(\eta^4).
\end{equation}
To integrate the remaining terms of Eq.(\ref{bl2}) I do 
the $u$-integration first and obtain
\begin{equation}
{\int}^{\infty}_{0}{dz} {\int}^{1}_{0}du\,\,
e^{-z\Lambda}
(\Lambda\coth z\sinh zu-\Lambda u) = \eta^2G(\eta) + \eta^2F({\lambda_\ell/ \eta})+\eta^2H(\eta,\lambda_\ell)-\eta I(\eta)
\label{udone}
\end{equation}
where I took $1-\lambda_\ell\simeq 1$ and $2-\lambda_\ell\simeq 2$ and introduced the four functions
\begin{equation}
F(x)={\int}^{\infty}_{0}{dz\over z^2}(z\coth z-1)
\left(2{e^{-zx}\over z}-{2\over z}+xe^{-zx} \right),
\label{F}
\end{equation}
\begin{equation}
G(x)={\int}^{\infty}_{0}{dz\over z^2}(z\coth z-1)
\left[{2\over z}(1-e^{-z/x})-{e^{-z/x}\over x}(2+{z\over x})\right] ,
\label{Gx}
\end{equation}
\begin{equation}
H(\eta,\lambda)={\int}^{\infty}_{0}{dz\over z^2}\coth ze^{-z\lambda/\eta}\left[{2\over (1-\eta^2)^2}-2+{\lambda\over\eta}{z\over 1-\eta^2}
-{\lambda\over\eta}z \right],
\label{Hx}
\end{equation}
and
\begin{equation}
I(x)={\int}^{\infty}_{0}{dz\over z^2}\coth ze^{-z/x}\left[{\sinh z +x^2\sinh z+2x\cosh z\over (1-x^2)^2}+{z\sinh z\over x-x^3}
+{z\cosh z\over 1-x^2} -2(x+z)-{z^2\over x}\right].
\label{Ix}
\end{equation}
A numerical evaluation of $F(x)$ shows that
\begin{equation}
F(x)\simeq-{2\over 3} \ln\left(1+6x \right)
\label{F2}
\end{equation}
with high accuracy. $G(x)$ is evaluated analytically for small $x$ by introducing a regulator $z^\epsilon$
and using the following series expansion of the hyperbolic cotangent
\begin{equation}
\coth z={1\over z} + 2z\sum_{n=1}^\infty {1\over (n\pi)^2+z^2}
\end{equation}
to obtain
\begin{equation}
G(x)=\lim_{\epsilon \rightarrow 0} \left[{2\over \pi^2}\pi^\epsilon\zeta (2-\epsilon) B\left({\epsilon\over 2},1-
{\epsilon\over 2}\right)-{2\over 3}x^\epsilon\Gamma(\epsilon)\right] -1.
\label{G2}
\end{equation}
Here $\zeta (2-\epsilon)$ is the Riemann zeta function, 
$B\left({\epsilon\over 2},1-{\epsilon\over 2}\right)$ is the Euler beta function and $\Gamma(\epsilon)$ is 
the Euler gamma function. After taking the limit we find
\begin{equation}
G(x)={2\over 3} \ln (\pi/x)+{2\over 3}\gamma_E-{4\over \pi^2}\zeta'(2) -1
\label{G3}
\end{equation}
where $\gamma_E=0.5772$ is the Euler-Mascheroni constant and $\zeta'(2)=-0.9375$ is the first derivative of the
Riemann zeta function. Last we evaluate $H(\eta,\lambda_\ell)$ and $I(\eta)$. After tedious integrations we find
\begin{equation}
\eta^2H(\eta,\lambda_\ell)={\cal{O}}(\eta^4),
\end{equation}
and 
\begin{equation}
\eta I(\eta)=-{\eta^2\over 6}+{\cal{O}}(\eta^4).
\end{equation}
Once we insert the expressions of $F$,$G$,$H$,$I$ and $J$ into Eq. (\ref{bl2}) we obtain the following
\begin{equation}
b_L=-{g^2\over 16\pi^2}\eta^2
\left[{2\over 3} \ln (\pi/\eta)+{2\over 3}\gamma_E-{4\over \pi^2}\zeta'(2) 
-{2\over 3} \ln\left(1+6{\lambda_\ell\over\eta} \right)+{1\over 6}
\right].
\label{bl4}
\end{equation}
In the case of a moderate field 
$\lambda_\ell/\eta\ll 1$ and $F(\lambda_\ell/\eta)=0$, 
and therefore Eq. (\ref{bl4}) gives the following
\begin{equation}
b_L={g^2\over 24\pi^2}\eta^2
\left[\ln \eta+{6\over \pi^2}\zeta'(2)-\ln \pi-\gamma_E -{1\over 4}
\right]
\end{equation}
where ${6\over \pi^2}\zeta'(2)-\log \pi-\gamma_E - {1\over 4}=-2.5418$ and it confirms the result of Ref. \cite{kuznetsov2}. 
It is very interesting to notice that $-\ln 6 -{3\over 4}=-2.5418={6\over \pi^2}\zeta'(2)-\log \pi-\gamma_E - {1\over 4}$, 
and this allows us to simplify significantly  Eq. (\ref{bl4}) and write it as
\begin{equation}
b_L={g^2\over 24\pi^2}\eta^2
\left[\ln \left( {\eta\over 6}+\lambda_\ell\right) -{3\over 4}
\right].
\label{b}
\end{equation}
This analytic expression is valid for $0\le eB \ll m^2_W$ and agrees, as I have already shown, with the 
known value of $b_L$ in the moderate field limit. It also agrees with the weak field value of $b_L$ obtained in  
Ref. \cite{kuznetsov2}, since in the weak field limit $\eta\ll\lambda_\ell$ and therefore $b_L={g^2\over 24\pi^2}\eta^2
\left(\ln \lambda_\ell -{3\over 4}
\right)$. An exact numerical 
computation of $b_L$ has also been done, and its results are reported in Figure 1, where 
the exact value of $b_L\times {24\pi^2\over g^2 \eta^2}$
is shown for  $0\le \eta \le 1$ and for the three neutrino species. 
Notice that $b_L$ diverges as $\eta$ approaches one.

All other coefficients of Eq. (\ref{coefficients}) have been evaluated and 
are listed below to leading order
\begin{equation}
a_L=-{g^2\over 48\pi^2}\eta^2
\left[\ln \left( {\sqrt{e}\over 6}\eta+\lambda_\ell\right) +{5\over 4}
\right],
\end{equation}
\begin{equation}
a_R={\epsilon_\nu\over 2}a_L,
\end{equation}
\begin{equation}
b_R={g^2\over 48\pi^2}\epsilon_\nu\eta^2
\left[\ln \left( {\eta\over 6}+\lambda_\ell\right) +{9\over 4}
\right],
\end{equation}
\begin{equation}
c_L={3g^2\over 32\pi^2}\eta,
\label{cl}
\end{equation}
\begin{equation}
c_R={g^2\over 64\pi^2}\epsilon_\nu\eta,
\label{cr}
\end{equation}
\begin{equation}
K_1={g^2\over 96\pi^2}\eta^2,
\end{equation}
\begin{equation}
K_2=-{g^2\over 64\pi^2}\lambda_\ell\eta(\ln\lambda_\ell+1).
\label{K2}
\end{equation}
The following 
\begin{equation}
{\int}^{\infty}_{0}{dz\over z^2}\left(\coth z - {1\over z}\right)\left(1-e^{-z\lambda/\eta}-
z{\lambda\over\eta}e^{-z\lambda/\eta}\right)\simeq{1\over 3}\ln\left(1+{6\over \sqrt{e}}
{\lambda\over \eta}\right)
\end{equation}
has been used when evaluating $a_L$.
The expression for $c_L$ agrees with that obtained in Ref. \cite{kuznetsov2}, 
the expressions for the other coefficients have not appeared in the literature.
\section{Discussion and conclusions}
\label{4}

I have calculated the neutrino self-energy in a magnetic field to one-loop order using the minimal extension of the Standard 
Model with massive Dirac neutrinos. My results for all the invariant coefficients of the
self-energy are valid for $0\le eB\ll m^2_W$ and for neutrino energies $0\le E \ll m_W$ and are 
reported in Eqs. (\ref{b}-\ref{K2}). These results make the distinction between weak and moderate fields obsolete
by providing simple 
and very accurate analytic expressions for all the coefficients, and
allow us to evaluate the neutrino self-energy around the critical magnetic field value 
$B_e=m^2_e/e\simeq 4.41 \times 10^{13}$ G,
where none of the previous calculations are valid. 
Some of the implications of these findings are discussed in the 
remaining part of this section.

The Dirac equation for left-handed massless neutrino is
\begin{equation}
\left[{\not\! p}-\Sigma(p)\right]L\nu_\ell(p)=0,
\end{equation}
and once we insert into it the expression (\ref{coefficients}) of the self-energy 
with $m_\nu=0$, we find the inverse neutrino propagator
\begin{equation}
S^{-1}_\nu={\not\! p}-a_L{\not\! p}-b_L(p\,\tilde\varphi\tilde\varphi\gamma)-c_L(p\,\tilde\varphi\gamma).
\end{equation}
By squaring the inverse propagator and setting it equal to zero we obtain the dispersion relation for massless neutrinos
\begin{equation}
(1-a_L)^2p^2-(2b_L-b^2_L-c^2_L-2b_La_L)\,p^2_{\parallel}=0
\end{equation}
which implies
\begin{equation}
{E\over |\bf{p}|}=1+\left(b_L+a_Lb_L-{b_L^2\over 2}-{c_L^2\over 2}\right)\sin^2\phi
\end{equation}
where $E$ and $\bf{p}$ are the neutrino energy and momentum and $\phi$ is the angle between $\bf{B}$ and $\bf{p}$. 
Since $a_Lb_L$, $b^2_L$ and $c^2_L$ are of higher order, in a perturbative sense, than $b_L$ they can be neglected and, using 
the coefficient $b_L$ obtained in this paper (\ref{b}), the dispersion 
relation can be written as
\begin{equation}
{E\over |\bf{p}|}=1-{G_F\over 3\sqrt{2}\pi^2}{(eB)^2\over m^2_W}
\left[\ln \left( {m^2_W\over {eB\over 6}+m^2_\ell}\right) +{3\over 4}
\right]\sin^2\phi,
\label{dispersion}
\end{equation}
where the term with the bracket is the magnetic field contribution.
Eq. (\ref{dispersion}) is valid for neutrino energies $E\ll m_W$ and applies to both $\nu_\ell$ and $\bar{\nu}_\ell$.

The neutrino dispersion relation in a magnetized medium has been studied in previous papers \cite{erdaskim,dolivo,elmfors}. In 
particular, a CP-symmetric medium with temperature $m_e\ll T\ll m_W$ and magnetic field $eB\le T^2$ was studied, because 
these conditions represent reasonably well the early universe plasma between the QCD phase transition and nucleosynthesis.
The following dispersion relation
\begin{equation}
{E\over |\bf{p}|}=1+{\sqrt{2}G_F\over 3}
\left[-{7\pi^2T^4\over 15}\left({1\over m^2_Z}+{2\over m^2_W}\right)+{T^2eB\over m^2_W}\cos\phi+
{(eB)^2\over 2\pi^2 m^2_W}\sin^2\phi\ln{T^2\over m^2_e}\right]
\label{dispersionmedium}
\end{equation}
was derived for $\nu_e$ and $\bar{\nu}_e$, where the first term is the pure plasma contribution \cite{notzold} and the other two terms 
are caused by the combined influence of plasma and magnetic field. Eq. (\ref{dispersionmedium}) was obtained
under the assumption that the magnetic field induced pure vacuum modification of the neutrino dispersion
relation was negligible. It turns out that the vacuum modification could be as 
large as the second or third term in Eq. (\ref{dispersionmedium}) and, 
once we include the vacuum modification presented in Eq. (\ref{dispersion}) of this paper, the dispersion relation for $\nu_e$ and $\bar{\nu}_e$ in a 
CP-symmetric plasma with magnetic field becomes
\begin{eqnarray}
{E\over |\bf{p}|}=&&1+{\sqrt{2}G_F\over 3}
\left[-{7\pi^2T^4\over 15}\left({1\over m^2_Z}+{2\over m^2_W}\right)+{T^2eB\over m^2_W}\cos\phi+
\right.
\nonumber \\
&&\left.+{(eB)^2\over 2\pi^2 m^2_W}\sin^2\phi\left(\ln{T^2\over m^2_e}-\ln{m^2_W\over {eB\over 6}+m^2_e}-{3\over 4}\right)\right]
\label{dispersionnew}
\end{eqnarray}
and is valid for $m_e\ll T\ll m_W$ and $0\le eB \le T^2$. When $eB\sim T^2$ the pure vacuum modification can be as large
as other terms of Eq. (\ref{dispersionnew}) and must be included in the neutrino dispersion relation. This dispersion relation
leads to an anisotropic neutrino index of refraction, causing neutrinos that move in the direction of the field to feel more the effect of the 
magnetized plasma (second term inside the bracket), while neutrinos moving perpendicularly to the field feel more the magnetization of the vacuum, 
similarly to what was found for strong fields in \cite{elizalde2}.

Another interesting application of my result, for the case of massive neutrinos, is the resonance enhancement of neutrino oscillations of the type 
$\nu_e\leftrightarrow \nu_{\mu,\tau}$. While it is well known that the medium alone can produce neutrino oscillation (i.e. the MSW effect),
I want to explore here the role of magnetic fields in neutrino oscillation enhancement. 
Conditions for resonance enhancement could be present, for example, inside an exploding supernova 
if a strong magnetic field is generated inside the exploding star, allowing for more energy transferred to the stellar matter by the $\nu_e$. 
The mixing angle $\theta_B$ in a magnetized medium is determined by the following relation
\begin{equation}
\sin^2 2\theta_B={\sin^2 2\theta\over{\left[\cos\theta\pm{2E(V_e-V_\ell)\over|\Delta m^2_\nu|}\right]^2}+\sin^2 2\theta}
\label{mixingangle}
\end{equation}
where $\Delta m^2_\nu$ and $\theta$ are the squared-mass splitting and vacuum mixing angle in the $\nu_e$, $\nu_\ell$ system and
$V_e$ and $V_\ell$ are the $\nu_e$ and $\nu_\ell$ effective potentials in the magnetized medium. The plus sign applies for $m_i< m_1$ 
and the minus sign for $m_i> m_1$, and $i=2,3$ for $\ell=\mu,\tau$. Even if the vacuum mixing angle $\theta$ is very small,
the mixing angle in magnetized matter is $\theta_B={\pi\over 4}$ if $m_i> m_1$ and $V_e-V_\ell$ is positive and satisfies the resonance condition
\begin{equation}
V_e-V_\ell={|\Delta m^2_\nu|\over 2E}\cos 2\theta.
\label{resonance}
\end{equation}
First I consider neutrinos propagating through a magnetized charged medium, where the chemical potential that displays the asymmetry
between particles and antiparticles is $\mu\ne 0$. In the case of a weak magnetic field $eB\ll\mu^2$, the pure magnetic field contribution to 
the effective potential is obtained immediately from Eq. (\ref{dispersion}) and, once we include it into $V_e-V_\ell$, 
Eq. (\ref{resonance}) becomes
\begin{equation}
\sqrt{2}G_FN_e\left(1+{eB\over 2m_e T}\cos \phi\right)-
{G_F\over 3\sqrt{2}\pi^2}{(eB)^2\over m^2_W}E\sin^2\phi
\ln \left( {eB+6m^2_\ell\over eB+6m^2_e}\right) 
={|\Delta m^2_\nu|\over 2E}\cos 2\theta
\label{resonance2}
\end{equation}
where $N_e$ is the electron number density and $T$ is the temperature of the medium. The first term of Eq. (\ref{resonance2})
is the thermal contribution of the magnetized medium to $V_e-V_\ell$ and was obtained in Ref.\cite{dolivo} for a 
nonrelativistic and nondegenerate electron gas. 

For a strong field $eB\gg\mu^2$ the thermal contribution of the magnetized medium to the neutrino self-energy was calculated
in Ref.\cite{elizalde2} and, once we include it into $V_e-V_\ell$, the resonance condition becomes
\begin{equation}
\sqrt{2}G_Fe^{-p^2_{\perp}/2eB}(N_e^0-N_{\bar{e}}^0)\left(1+\cos \phi\right)-
{G_F\over 3\sqrt{2}\pi^2}{(eB)^2\over m^2_W}E\sin^2\phi
\ln \left( {eB+6m^2_\ell\over eB+6m^2_e}\right) 
={|\Delta m^2_\nu|\over 2E}\cos 2\theta
\label{resonance3}
\end{equation}
where $N_e^0$ and $N_{\bar{e}}^0$ are the electron and positron number densities in the lowest Landau level.
Using a supernova core density of $10^{17}$ Kg/m$^3$
we have $\sqrt{2}G_FN_e\simeq 3.8$ eV and, neglecting the neutrino masses and for $E=10$ MeV, I find that a magnetic field strength 
$B\sim 10^{22}$ G is required for 
the resonance transition $\nu_e\leftrightarrow \nu_{\tau}$ to occur inside an exploding supernova, far exceeding the magnetic field believed 
to exist inside the supernova. 

The case of a magnetized neutral medium ($\mu=0$) should also be explored. It was shown in Ref.\cite{ferrer} that, 
in the case of a strong magnetic field $eB\gg T^2$, pure magnetic neutrino oscillations are possible in a $CP$-symmetric magnetized medium.
I find that, for $T^2\gg eB$ and $T^2\gg m^2_\mu$, the resonance condition for pure magnetic $\nu_e\leftrightarrow \nu_{\mu}$ oscillations is
\begin{equation}
{G_F\over 3\sqrt{2}\pi^2}{(eB)^2\over m^2_W}E\sin^2\phi\left[\ln{m^2_\mu\over m^2_e}-
\ln \left( {eB+6m^2_\mu\over eB+6m^2_e}\right)\right]
={|\Delta m^2_{21}|\over 2E}\cos 2\theta_{12}.
\label{resonance5}
\end{equation}
Since $\Delta m^2_{21}=8\times 10^{-5}$ eV$^2$ and $\cos 2\theta_{12}=0.51$ \cite{pdg}, Eq. (\ref{resonance5}) 
shows that for a neutrino energy $E=1$ GeV resonance will occur when $B\sim 2m^2_e/e=8.8 \times 10^{13}$ G. This finding can have interesting
implications for cosmology, showing that resonant neutrino flavor oscillations could be caused by a moderate magnetic field in the primeval plasma.

The magnetic field induced resonance transition $\nu_{\mu}\leftrightarrow \nu_{\tau}$ in a magnetized charged medium 
should also be investigated in this context and, for 
this transition to occur, the resonance condition is 
\begin{equation}
{\Delta m^2_{32}\over 2E}\cos 2\theta_{23}+
{G_F\over 3\sqrt{2}\pi^2}{(eB)^2\over m^2_W}(E\sin^2\phi)
\ln \left( {eB+6m^2_\tau\over eB+6m^2_\mu}\right)=0,
\label{resonance4}
\end{equation}
since the charged current contribution to the effective potential of $\nu_{\mu}$ and $\nu_{\tau}$ in a medium is absent. Eq. (\ref{resonance4}) shows
that resonance could occur only if  $\Delta m^2_{32}$ is negative, which is not ruled out \cite{pdg}.
For $E=10$ MeV and $\Delta m^2_{32}\simeq -2.5\times 10^{-3}$ eV$^2$ \cite{pdg} I find that a magnetic field $B\simeq 8.5\times 10^{2}
\times\sqrt{\cos2\theta_{23}}B_e$ is necessary for 
resonance to occur. Since the accepted value of the mixing angle is $37^0\le\theta_{23}\le 45^0$ \cite{pdg} and $B_e\simeq 4.41 \times 10^{13}$ G, 
we will have resonance conditions for $B\le 2\times 10^{16}$ G, a magnetic field strength that could exist inside magnetars.

One more case where the results obtained in this paper are relevant, is the calculation of the magnetic moment $\mu_{\nu_\ell}$ 
of ${\nu_\ell}$ within the minimally extended standard model of the electroweak interactions containing an $SU(2)$-singlet 
right-handed neutrino. The neutrino magnetic moment can be written in 
terms of the self-energy coefficients of Eq. (\ref{coefficients}) as \cite{kuznetsov3}
\begin{equation}
\mu_{\nu_\ell}={m_\nu\over 2 B}\left(c_L-c_R+4K_2\right)
\label{magneticmoment}
\end{equation}
where 
\begin{equation}
c_L={g^2\over 16\pi^2}
{\int}^{\infty}_{0}{dz}
{\int}^{1}_{0}du\,\,
{e^{-z\Lambda}\over\sinh z}u\left[\sinh(z+zu)-{\lambda_\ell\over 2}\sinh(z-zu)\right]
\label{cl2}
\end{equation}
\begin{equation}
K_2={g^2\over 16\pi^2}\left({\lambda_\ell\over 2}\right)
{\int}^{\infty}_{0}{dz}
{\int}^{1}_{0}du\,\,e^{-z\Lambda}
{\sinh(z-zu)\over\sinh z}
\label{k22}
\end{equation}
and $c_R$ is given by Eq. (\ref{cr}). Once we retain all the sub-leading $\lambda_\ell$-corrections to $c_L$ and $K_2$, we have
\begin{equation}
c_L={g^2\eta\over 32\pi^2}{1\over(1-\lambda_\ell)^3}
\left(3-{17\over 2}\lambda_\ell+5\lambda_\ell^2+{\lambda_\ell\over2}+2\lambda_\ell\ln\lambda_\ell-5\lambda_\ell^2\ln\lambda_\ell
\right)+{\cal{O}}(\eta^3)
\label{cl3}
\end{equation}
\begin{equation}
K_2
=-{g^2\eta\over 32\pi^2}{1\over(1-\lambda_\ell)^2}\left({\lambda_\ell\over 2}\right)
\left(\ln\lambda_\ell+1-\lambda_\ell\right)+{\cal{O}}(\eta^3)
\label{k23}
\end{equation}
and, for $m_\nu\ll m_\ell$, we find
\begin{equation}
\mu_{\nu_\ell}=\mu_{\nu_\ell}^{(0)}{1\over(1-\lambda_\ell)^3}\left(1-{7\over 2}\lambda_\ell+3\lambda_\ell^2
-\lambda_\ell^2\ln\lambda_\ell-{1\over 2}\lambda_\ell^3\right),
\label{magneticmoment2}
\end{equation}
where the leading term $\mu_{\nu_\ell}^{(0)}$ is \cite{lee,fujikawa}
\begin{equation}
\mu_{\nu_\ell}^{(0)}={3eG_Fm_\nu\over 8\pi^2\sqrt{2}}.
\label{magneticmoment3}
\end{equation}
The neutrino magnetic moment of Eq. (\ref{magneticmoment2}) agrees with the results of Refs. \cite{CabralRosetti,Dvornikov}, 
obtained by different methods.
\begin {acknowledgements} 
A. Erdas wishes to thank Marcello Lissia for helpful discussions,
the Department of Physics of the University of Cagliari and the I.N.F.N. Sezione di Cagliari
for their continued support and the High Energy Theory Group of the Johns Hopkins University for the
hospitality extended to him during his several visits.
\end {acknowledgements} 

\begin{figure}[fg1]
\psfig{figure=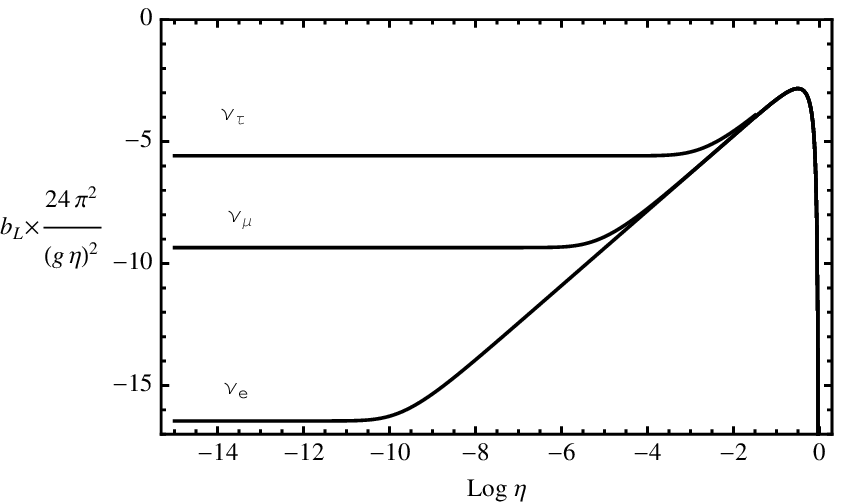
,height=8.25cm
,width=13.5cm
}\caption[fc1]{
Exact value of $b_L\times {24\pi^2\over g^2 \eta^2}$ for $10^{-15}\le {eB\over m^2_W}\le 10^0$ 
for the three neutrino species. For $-\infty <\log\eta\le -1$ the analytic expression of $b_L$ obtained 
in Eq. (\ref{b}) of this paper
is in within 0.27~\% or less of the exact numerical value.
\label{fig1}
           }
\end{figure}

\end{document}